\documentclass[prb,aps,epsf,showpacs,twocolumn,preprintnumbers,amsmath,amssymb]{revtex4}
\usepackage{graphicx}
\usepackage{dcolumn}
\usepackage{bm}
\begin{document}


\title{Superflow in Solid $^4$He} 
\author{ Wayne M. Saslow} 
\email{wsaslow@tamu.edu}
\affiliation{ Department of Physics, Texas A\&M University, College Station, 
TX 77843-4242}
\date{\today}

\begin{abstract}
Kim and Chan have recently observed Non-Classical Rotational Inertia (NCRI) for solid $^4$He in Vycor glass, gold film, and bulk.  Their low $T$ value of the superfluid fraction, $\rho_{s}/\rho\approx0.015$, is consistent with what is known of the atomic delocalization in this quantum solid.  By including a lattice mass density $\rho_{L}$ distinct from the normal fluid density $\rho_{n}$, we argue that $\rho_{s}(T)\approx\rho_{s}(0)-\rho_{n}(T)$, and we develop a model for the normal fluid density $\rho_{n}$ with contributions from longitudinal phonons and ``defectons'' (which dominate).   The Bose-Einstein Condensation (BEC) and macroscopic phase inferred from NCRI implies quantum vortex lines and quantum vortex rings, which may explain the unusually low critical velocity and certain hysteretic phenomena. 

\end{abstract}



\maketitle

{\bf Introduction.} 
In 1969, Andreev and Lifshitz proposed that solid $^4$He could display unusual lattice flow properties if the crystal vacancies and defects become mobile enough for quantum diffusion to become more significant than thermal diffusion.\cite{A&L}  They suggested that external objects placed within the solid could become unusually mobile, and they developed a two-fluid hydrodynamics for the system.  In 1970, motivated by Monte Carlo studies of the wavefunction for solid $^4$He, Chester suggested that solid $^4$He might exist in a periodic state that is subject to Bose-Einstein Condensation (BEC) and, further, that an alternative way to obtain BEC would be for the system to have vacancies in the ground state.\cite{Chester}  Leggett shortly proposed that solid $^4$He might display Non-Classical Rotational Inertia (NCRI), where under rotation the angular momentum would be less than for a classical state.\cite{Leggett1}  In the rotating frame the lattice is at rest, but there is mass flow due to gradients in the phase $\phi$ of the wavefunction, which one can call ``phase flow'' to distinguish it from lattice flow.  Ref.~\onlinecite{Leggett1}  developed a theoretical expression for the excess energy caused by rotation.  It also indicated how to obtain an upper limit -- and for BEC perhaps the exact value -- for the superfluid fraction $\rho_{s}/\rho$ at temperature $T=0$, where $\rho_{s}$ is the superfluid mass density and $\rho$ is the total mass density.  Estimates based on analogies to tunnelling in solid $^3$He led to values of the superfluid fraction from 0.01\%\cite{Leggett1} to 0.0001\%,\cite{Guyer} but actual calculations based on Leggett's theory led to values from 5\% to 30\%.\cite{FernandezPuma,Saslow1}  

Since these suggestions there have been many searches for ``supersolid'' behavior.\cite{Meisel}  To date, no experiment has found evidence for lattice flow.\cite{Andreev,Greywall}  However, using the torsion oscillator technique, Kim and Chan recently have observed NCRI for solid $^4$He in Vycor glass.\cite{KimChan}  Even more recently they have found NCRI for solid $^4$He in porous Au and in bulk.\cite{KimChan2}  They find that the zero temperature value for $\rho_{s}/\rho$ is on the order of 1-2\%.  They also find that the transition from the supersolid to the ordinary solid (which does not display NCRI) is around 0.2~K.  

The present work: (1) obtains converged values for the superfluid fraction at $T=0$ in terms of the localization of the density about the lattice sites, and using the experimental superfluid fraction finds reasonable agreement with independent measures of that localization; (2) presents a theory for the temperature-dependence of $\rho_{s}$ in terms of a lattice mass density $\rho_{L}$ and the temperature-dependent normal fluid density $\rho_{n}$, which qualitatively explains how the transition temperature can be so much lower than in liquid $^3$He; (3) proposes simple forms for the phonon and ``defecton'' contributions to $\rho_{n}$, which leads to semi-quantitative agreement with experiment and with the observed transition temperature $T_{c}$; (4) proposes that quantum vortex lines can occur, and that they have low enough energies to explain the factor of 700 lower critical velocities observed in the  supersolid as opposed to the superfluid;\cite{KimChan} (5) proposes that quantum vortex rings can occur, and that they may explain the hysteresis observed on cooling but not on heating; (6) proposes that adding $^3$He not only lowers the superfluid fraction by increasing the normal fluid component, but also may raise $T_{c}$  by providing pinning sites that prevent vortex binding and unbinding, and thus inhibit collective processes responsible for the transition; (7) suggests that quantum vortex rings in the supersolid might be produced, as in the superfluid, with ion sources. 

{\bf Superfluid Density at $T=0$.}
Consider the flow energy $E=(1/2)\int\rho({\vec{r}})v^{2}_{s}({\vec{r}})d\vec{r}$, where 
$\rho({\vec{r}})$ and $\vec{v}_{s}({\vec{r}})=(\hbar/m_{4})\vec{\nabla}\phi(\vec{r})$ are the local values of the mass density and superfluid velocity.\cite{Leggett1}  (Here $m_{4}$ is the bare $^4$He mass.)  Minimization with respect to the phase $\phi$ leads to the superflow condition $\vec{\nabla}\cdot\big(\rho(\vec{r})\vec{v}_{s}(\vec{r})\big)=0$.\cite{Saslow1}  From a knowledge of $\rho({\vec{r}})$ and the spatial average superfluid velocity $\vec{v}_{s}$, one can determine the full superfluid velocity profile $\vec{v}_{s}({\vec{r}})$, and from that the flow energy $E$.  The superfluid density then follows on setting $E/V=(1/2)\rho_{s}{v}^{2}_{s}$, where $V$ is the system volume.  Physically, the minimization condition causes phase flow (from $\vec{v}_{s}(\vec{r})$) to occur preferentially in regions of low density, and the more localized the $\rho(\vec{r})$, the more suppressed is  $\rho_{s}/\rho$ from unity.  Note that only $\rho(\vec{r})$ is defined on a microscopic level; its spatially average value, the total mass density $\rho$, as well as $\rho_{s}$, are defined only on the macroscopic level. 

Ref.~\onlinecite{Saslow1} computed $\rho_{s}/\rho$ as a function of $b/a$, for a $\rho({\vec{r}})$ given as a sum over Gaussians of width $b$ placed on the sites of a fictitious face-centered-cubic lattice of solid $^4$He, with lattice constant $a$.   In Refs.~\onlinecite{KimChan}and \onlinecite{KimChan2} the system was likely a hexagonal close-packed crystal or polycrystal.  Since the hcp and fcc lattices have the same local packing, this use of an fcc lattice is appropriate.\cite{CouplingNote}

At the lower values of $b/a$ the results of Ref.~\onlinecite{Saslow1} were only an upper bound, because the computer then available did not have the capability to solve for the superflow pattern accurately.\cite{Saslow1}  Current computers permit convergent calculations.  Figure~1 gives values for $\rho_{s}/\rho$ for the full range of $b/a$.  It gives the expected result that $\rho_{s}/\rho$ approaches unity for a delocalized system (large $b/a$); for $b/a=0.30$, $\rho_{s}/\rho\approx0.9860$.  It also gives the expected result that $\rho_{s}/\rho$ approaches zero for a localized system (small $b/a$); for $b/a=0.10$, $\rho_{s}/\rho\approx0.0009$.  For $0.10\le b/a\le 0.20$, $\rho_{s}/\rho$ is very sensitive to $b/a$, as shown by the inset.   Over 6000 plane waves were needed to obtain $\rho_{s}/\rho$ at the smaller values of $b/a$.  Although $\rho_{s}$ can be a tensor for a general lattice, it is diagonal for a cubic lattice; for the fcc lattice considered her, on convergence $\rho_{s}$ was independent of the direction of superflow. 


\begin{figure}
\centering
  \includegraphics[angle=0,width=3.8in]{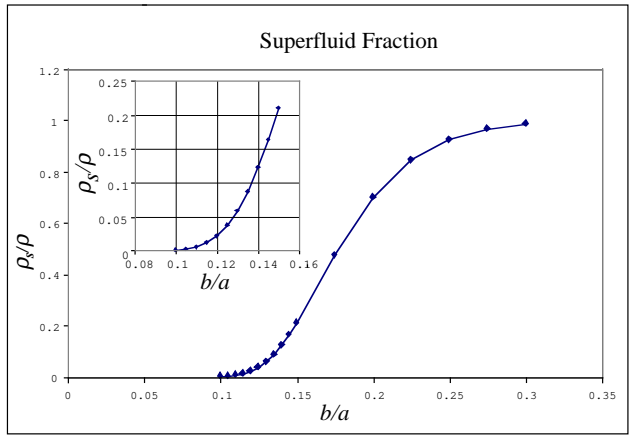}
  \caption{\label{fig:epsart} Superfluid fraction $\rho_s/\rho$ {\it vs} $b/a$.}
  \label{fig1}
\vskip-0.1in
\end{figure}

The exchange considerations of Ref.~\onlinecite{Guyer} give $b/a$ near 0.125.  The inset  yields that $\rho_{s}/\rho=0.022$ for $b/a=0.125$.  Considering the sensitivity of $\rho_s/\rho$ to $b/a$, this is in good agreement with the experimental value $\rho_{s}/\rho\approx0.015$.  Because $\rho_{s}/\rho$ is so sensitive to $b/a$, it would be desireable to have a more accurate $\rho(\vec{r})$ from calculations on solid $^4$He using wavefunctions possessing BEC.\cite{LowyWoo,CepChesKal}  

If the $\rho(\vec{r})$ for finite $T$ is employed in the above procedure, one obtains a quantity 
$\rho_{s}^{*}(T)$, with $\rho_{s}^{*}(T=0)=\rho_{s}(T=0)$, whose significance we discuss next. 

{\bf Superfluid Density at Finite $T$; Lattice Density.}
Following Ref.~\onlinecite{A&L}, $\rho_{s}$ and $\rho_{n}$ are (implicitly) tensors.  In Ref.\onlinecite{Saslow2} the author attempted to extend the hydrodynamics of Ref.\onlinecite{A&L} to include the fact that superflow is defined relative to the lattice.  That work, however, did not go far enough because it did not explicitly include the lattice velocity $\vec{v}_{L}$ or the lattice mass density $\rho_{L}$ in the momentum density.  To incorporate this dependence of the superflow on motion relative to the lattice, we take the total momentum density to have the form 
\begin{equation}
\vec{g}=\rho_{s}\vec{v}_{s}+\rho_{n}\vec{v}_{n}+\rho_{L}\vec{v}_{L} 
	=\rho_{s}(\vec{v}_{s}-\vec{v}_{L})+\rho_{n}(\vec{v}_{n}-\vec{v}_{L})+ \rho\vec{v}_{L},
\label{S-g}
\end{equation}
where $\vec{v}_{L}=d\vec{u}/dt$ is the lattice velocity and $\vec{u}$ is the lattice displacement relative to a fiducial state.  Here the (net) average mass density satisfies 
\begin{equation}
\rho=\rho_{s}+\rho_{n}+\rho_{L}, 
\label{rho-tot}
\end{equation}
and $\vec{g}\rightarrow\vec{g}+\rho\vec{v}$ under a Galilean boost of all three velocities by $\vec{v}$.  Thus we take the system to initially possess three independent mass densities and velocities, rather than the usual two independent mass densities and velocities.  Although the macroscopic dynamical equations\cite{A&L,Saslow1,Liu} predict that $\vec{v}_{n}\approx\vec{v}_{L}$, to develop a theory for $\rho_{s}$ and $\rho_{n}$ it is essential at the outset not to make this simplification.  On applying the macroscopic dynamical equations, the normal modes become essentially the same as in Ref.\onlinecite{A&L}, subject to the comments of Liu.\cite{Liu,Note1} 

With $\rho_{s}^{*}(T)$ of the previous section, we identify the lattice density $\rho_{L}(T)=\rho(T)-\rho_{s}^{*}(T)$.  Then, by (\ref{rho-tot}),
\begin{equation}
\rho_{s}(T)=\rho(T)-\rho_{L}(T)-\rho_{n}(T)=\rho_{s}^{*}(T)-\rho_{n}(T), 
\label{rho_s}
\end{equation}
rather than $\rho_{s}(T)=\rho(T)-\rho_{n}(T)$, as holds for bulk superfluid.  Hence, if $\rho_{s}(T_{c})=0$ defines $T_{c}$, and $\rho_{s}/\rho=0.015$ at $T=0$, in the supersolid one needs only a normal fluid fraction of 0.015 (rather than unity in the superfluid) to suppress superflow.  This would explain why  $T_{c}$  for bulk solid $^4$He is only about 0.2~K, whereas it is 2.176~K for bulk liquid $^4$He.  In practice, at temperatures low enough for NCRI to be observed, $\rho(\vec{r})$ may be essentially independent of temperature, so that we may be able to employ  $\rho_{s}^{*}(T)\approx\rho_{s}^{*}(0)=\rho_{s}(0)$ in (\ref{rho_s}).  Like $\rho$ and $\rho_{s}$, $\rho_{n}$ and $\rho_{L}$ are defined only on the macroscopic level.  

By contrast, consider liquid $^4$He in a superleak.  For pores wide enough that the transition is not suppressed, the pore superfluid density $\rho^{(p)}_{s}$ is geometrically suppressed but otherwise takes on bulk values, so $\rho^{(p)}_{s}=g{\cal P}\rho_{s}$, where $g<1$ is a geometrical constant reflecting the tortuous flow pattern in the superleak, ${\cal P}$ is the porosity, and $\rho_{s}=\rho-\rho_{n}$ is the temperature-dependent bulk superfluid density (so that the transition temperature takes on the bulk value).\cite{new4thref}  On the other hand, for pores narrow enough to suppress $T_{c}$, $\rho_{s}^{(p)}$ should reflect both the geometrical constraints of the geometry via $g$ and ${\cal P}$, as well as a factor that differs from $\rho_{s}$, to reflect the microscopic geometrical suppression of the overall superfluid density (e.g., due to suppression or modification of the order parameter).\cite{new4thref}   Neither of these types of superleak has the same behavior as proposed in  (\ref{rho_s}) for solid $^4$He.  

{\bf Normal Fluid Density.}
As a first pass at a concrete expression for $\rho_{n}$, we ignore the distinction between momentum and crystal momentum; a proper theory thus must be more complex than in Ref.~\onlinecite{Landau}.\cite{Note3}  We expect two contributions to $\rho_{n}$: first, from phonons (here we consider only longitudinal phonons, with characteristic velocity $v_{L}\approx500$~m/s); and second from defect excitations, or ``defectons'' (perhaps ``vacancions'', perhaps ``interstitial-ons'', perhaps excitons consisting of coupled motions of both).  We take their excitation spectrum to have the form\cite{A&L,Hetherington,Guyer2,GuyerRichZane} 
\begin{equation}
\varepsilon=\Delta_{4}+\frac{p^2}{2\mu_{4}},
\label{excitation}
\end{equation}
where $\Delta_{4}$ is the minimum energy of excitation, $p$ is crystal momentum, and $\mu_{4}$ is their effective mass.  From Landau\cite{Landau, L&L-SP1} these two types of boson excitations give
\begin{equation}
\rho_{n}=\frac{ 2\pi^{2}k_{B}^{4}T^{4} }{ 45\hbar^{3}v_{L}^{5} }
+\mu_{4}\bigl( \frac{ \mu_{4}k_{B}T }{ 2\pi\hbar^{2} } \bigr)^{3/2} \exp{ (-\Delta_{4}/k_{B}T) }.
\label{rho_n-theory}
\end{equation}

The phonon contribution to $\rho_{n}$ is $10^{-8}$ of $\rho$ for $T=0.2$~K, and thus is negligible; the exponential term due to ``defectons'' dominates.  $\Delta_{4}=2$~K and $\mu_{4}$ of about 60\% of $m_{4}$ give a qualitative fit to the data, thus supporting this view of the origin of the normal fluid, as well as what is currently known about defect excitations.  Taking $\Delta_{4}=1$~K\cite{Goodkind2} requires a much lower effective mass to give a similar quality fit the data.  Although (\ref{rho_n-theory}) explains  the sudden onset of $\rho_{n}$, it does not explain the long tail observed at higher temperatures.\cite{KimChan,KimChan2} 

The addition to solid $^4$He of a small concentration $c$ of $^3$He should change $\rho$ and $\rho_{L}$ in proportion to $c$, or  
\begin{equation}
\rho_{s}(T=0,c)=\rho_{s}(T=0)-\alpha\rho c,
\label{rho_s-3}
\end{equation}
where $\alpha$ is a dimensionless constant.  Moreover, at low $c$, $\rho_{n}$ likely will be suppressed in proportion to $c$, because the system has less $^4$He to excite.  The data have this qualitative behavior.\cite{KimChan,KimChan2} 

{\bf Vortex Lines and Vortex Rings.}
One striking difference between superflow in liquid and solid $^4$He is that the critical velocity $v_{c}$ is much lower for the solid ($\approx300\mu$m/s, as opposed to $\approx200$ mm/s).\cite{KimChan}  In torsion oscillator experiments, during an oscillation the velocity amplitude ranges from zero to a maximum value $v_{m}$.  If $v_{m}<v_{c}$, there is no degradation of the superflow, but if $v_{m}>v_{c}$, there is degradation during that part of the oscillation where $v>v_{c}$.  The observed $v_{c}$ is on the order of what one would obtain for a single unit of circulation.\cite{KimChan2}  

The observation of NCRI implies BEC, making the phase $\phi$ of the wavefunction a well-defined quantity and permitting the superflow pattern calculations described above.  Another aspect of having phase as a well-defined quantity is that we can expect other excitations involving the phase.  In particular, there should be quantum vortex lines and quantum vortex rings.\cite{Onsager, Feynman}  This is relevant to the low observed values of the critical velocity. 

Just as the non-uniform density of the solid makes the superflow pattern non-uniform, thereby decreasing the average superfluid density (as in Figure 1), so the non-uniform density of the solid makes it easier for vortices to enter the system and destroy the superflow.  This is because there are low-density regions in the solid (i.e. between the crystal sites), at which quantum vortices will preferentially nucleate.\cite{Fetter,SaddChesReat}  This preference occurs because the largest energy cost for a quantum vortex in a finite system is associated with the formation of its core, where the local density tends to go to zero.  If the local density is already suppressed by 90\% between the crystal sites, it costs the solid much less energy than the liquid to further reduce the local density to zero.  Without calculating in detail the vortex flow velocity field (by analogy to that of Ref.\onlinecite{Saslow1} for bulk flow), it is not possible to say how much less energy would be required to create a vortex line in supersolid $^4$He relative to superfluid $^4$He.\cite{VortexNote}  However, since the energy to create a vortex line is proportional to $\rho_{s}$, which at low temperature is suppressed from $\rho$ by almost a factor of one hundred, and since the critical velocity is proportional to the energy to create a vortex line,\cite{Feynman} the  suppression of $\rho_{s}$ by non-uniform superflow provides a mechanism for understanding most of the factor of 700 difference between the critical velocities in the supersolid and the superfluid.\cite{KimChan2}  Presently, spread in the data due to nonuniformity of the $^4$He crystals prevents a meaningful comparison between $\rho_{s}$ and $v_{c}$.\cite{Chan, VortexSpreadNote, NormalFluidNote}  

Indirect evidence for vorticity (lines and rings) comes from the asymmetry under changes of temperature of the equilibration time $\tau$ between torsion oscillator runs.  On warming to $T$, the $\tau$ to the next run is the same as between consecutive runs at $T$; but on cooling, $\tau$ is larger than between consecutive runs at $T$; moreover, this time lengthens with decreasing temperature.\cite{KimChan}  This is consistent with thermal vorticity being produced on warming, but on cooling excess thermal vorticity requiring additional time to equilibrate at the lower temperature.  Some of this vorticity may be  induced by the oscillation itself, in which case small amplitude oscillations will equilibrate faster on cooling.  (Some equilibration of the system is due simply to the large $Q\approx10^{6}$ of the oscillator, which coupled with a period $P\approx10^{-3}$~s, gives a 1000~s ring-down time.\cite{KimChan}) 

Although quantum vortices are well known to occur in superconductors,\cite{Abrikosov} which are solids, the vortices occur in the phase $\phi$ of the wavefunction for the electron fluid within a solid structure determined by atoms.  The present case is more extreme, for the vortices occur in the $^4$He, which is also responsible for the solid structure: supersolid $^4$He may be thought of as a localized fluid, rather than merely as a solid. 

{\bf Effect of $^3$He on the Transition Temperature.}
If quantum vortex lines and quantum vortex rings are responsible for the details of the transition from supersolid to ordinary solid at about 0.2~K, then the addition of $^3$He will provide pinning centers that suppress vortex motion and prevent vortex proliferation.  This should make it more difficult to destroy superfluidity, and thus raise the transition temperature, as observed.\cite{KimChan,KimChan2}  Such enhancement is in contrast to what happens at low temperatures, where the $^3$He suppresses the superfluid fraction.  The data also shows a long, drawn-out persistence of superfluidity at higher temperatures;\cite{KimChan, KimChan2} this might be a dynamical effect,\cite{Noteinhomo} disappearing for an oscillator with a significantly longer period, for which the vortices would have more time to escape the pinning centers.\cite{AHNS}  

{\bf Final Remarks.}
Atoms in optical lattices also are candidates for supersolid behavior.\cite{Greiner,Meyerovich,CornWiem,Leggett2}  Their low moment of inertia and the imposed symmetry-breaking optical lattice may make it more appropriate to study their normal modes rather than their non-classical moment of inertia.  Nevertheless, the transition from the ($\rho_{s}=0$) Mott insulating state to the ($\rho_{s}\ne0$) BEC state, which can in principle be produced by decreasing the strength of the laser field producing the lattice, should provide a dramatic change in the moment of inertia.  Likewise, in the BEC state, increasing the laser field strength should further localize the system, and decrease $\rho_{s}$.  

Although the unusual solid flow (as opposed to ``phase flow'') effects suggested by Ref.~\onlinecite{A&L} have not been observed, with $\rho_{n}\rightarrow\rho_{n}+\rho_{L}$\cite{Note1} this work may correctly  provide the hydrodynamics of supersolids.  

It would be of great interest to find more direct evidence for the vortex excitations of the macroscopic phase $\phi$, whose existence is inferred from the observation of NCRI.\cite{KimChan,KimChan2}  Perhaps vortex rings can be observed using a method similar to that of Rayfield and Reif.\cite{RayReif}

{\bf Acknowledgements.}
I would like to thank M. H. W. Chan for valuable discussions of his results, and for making available a preprint of Ref.\onlinecite{KimChan2}.   This work was supported in part by the Department of Energy, through DOE Grant No. DE-FG03-96ER45598.

{}


\begin{thebibliography}{}
\bibitem{CouplingNote} In the samples of Ref.~\onlinecite{KimChan2}, coupling at the interfaces between different crystallites may be incomplete,  which would decrease the experimental value of $\rho_{s}$.
\bibitem{VortexNote} Far from the vortex core, the flow is nearly uniform over a unit cell, where the flow pattern and energy density should be very like that discussed in Refs.~\onlinecite{Leggett1} and  \onlinecite{Saslow1}.  Near the vortex core the flow pattern can be rather different.  In the superfluid, the energy per unit length $L$ of a vortex line at the center of a cylinder of radius $R$ is given by $E/L=\pi\rho_{s}(\hbar/m_{4})^{2}\ln(R/a)$, where $a$ is a length on the order of the size of the vortex core.  In the supersolid this same form will hold, with the reduced value of $\rho_{s}$.  In addition, because the energy of nucleation (which involves the energy at small distances) is expected to be relatively small independently of the smaller value of $\rho_{s}$, we expect that $a$ for the solid will be relatively large.
\bibitem{VortexSpreadNote} Because the experimental systems are expected to be polycrystalline, the surfaces will be nonuniform, with nonuniform energies for vortex nucleation.
\bibitem{NormalFluidNote} Polycrystals may have poor interfacial couplings that suppress $\rho_{s}^{*}$, enhance $\rho_{L}$, and suppress $\rho_{s}$.  However, the excitation-dependent $\rho_{n}$ may be  sample-independent.
\bibitem{Noteinhomo} If due to inhomogeneities, it may disappear for a single crystal.
\end{thebibliography}
\end{document}